\documentclass{elsart1p}

\usepackage{epsfig}
\usepackage{amsfonts}
\usepackage{bbm}

\newcommand{\tb}{} 
\newcommand{\tr}{} 

\newcommand{\mD}{m_\rmii{\hspace*{-0.0mm}D}}
\newcommand{\rmO}{{\mathcal{O}}}

\renewcommand{\vec}[1]{{\bf #1}}
\newcommand{\nwc}{\newcommand}
\nwc{\nl}  {\newline}
\nwc{\be}  {\begin{displaymath}}
\nwc{\ee}  {\end{displaymath}}
\nwc{\bmu} {\bar{\mu}}
\nwc{\ba}  {\begin{eqnarray*}}
\nwc{\ea}  {\end{eqnarray*}}
\nwc{\bc}  {\begin{center}}
\nwc{\ec}  {\end{center}}
\nwc{\bi}  {\begin{itemize}}
\nwc{\ei}  {\end{itemize}}
\nwc{\nn}  {\nonumber\\}
\nwc{\Tr}  {\mathop{\rm Tr}}
\nwc{\Det} {\mathop{\mbox{Det}}}
\nwc{\diag} {\mathop{\mbox{diag}}}
\nwc{\re}  {\mathop{\rm Re}}
\nwc{\im}  {\mathop{\rm Im}}
\nwc{\Hc}  {\mathop{\rm H.c.}}
\nwc{\la}[1]{\label{#1}}
\nwc{\rmi}[1]{{\! \mbox{\scriptsize #1}}}
\nwc{\rmii}[1]{{\mbox{\tiny\rm{#1}}}}
\nwc{\nr}[1]{(\ref{#1})}
\nwc{\fr}[2]{{\frac{#1}{#2}}}
\nwc{\msbar}{\overline{\mbox{\rm MS}}}
\nwc{\lambdamsbar}{\Lambda_{\overline{\rm MS}}}
\newcommand{\tinymsbar}{{\overline{\mbox{\tiny\rm{MS}}}}}

\newcommand{\Nc}{N_{\rm c}}

\newcommand{\fig}{Fig.~}
\newcommand{\se}{Sec.~}

\def\lsi{\raise0.3ex\hbox{$<$\kern-0.75em\raise-1.1ex\hbox{$\sim$}}}
\def\gsi{\raise0.3ex\hbox{$>$\kern-0.75em\raise-1.1ex\hbox{$\sim$}}}
\nwc{\lsim}{\mathop{\lsi}}
\nwc{\gsim}{\mathop{\gsi}}

\newcommand{\unit}{{\mathbbm{1}}} 
\newcommand{\hide}[1]{ }

\begin{document}

\begin{frontmatter}




\hfill{BI-TP 2008/24} \\
\hfill{arXiv:0810.1112}

\vspace*{-1.5cm}

\title{How to compute the thermal quarkonium \\
spectral function from first principles?}


\author{M.~Laine}

\address{Faculty of Physics, University of Bielefeld, 
D-33501 Bielefeld, Germany}

\begin{abstract}
In the limit of a high temperature $T$ and a large quark-mass $M$,   
implying a small gauge coupling $g$, the heavy 
quark contribution to the spectral function of the electromagnetic current 
can be computed systematically in the weak-coupling expansion. We argue
that the scale hierarchy relevant for addressing 
the disappearance (``melting'') of the resonance peak from 
the spectral function reads 
$M \gg T > g^2 M > gT \gg g^4 M$, 
and review how the heavy scales can be integrated out one-by-one, 
to construct a set of effective field theories describing the low-energy
dynamics. The parametric behaviour of the melting temperature in the
weak-coupling limit is specified. 
\end{abstract}

\end{frontmatter}

%
\section{Introduction}

Given possible applications in heavy ion collision experiments~\cite{ms}, 
the last 20 years have seen a huge amount of phenomenological 
work on the properties of heavy quarkonium at temperatures just above 
the deconfinement crossover (for a recent review, see ref.~\cite{rbc}).
During the last couple of years, these phenomenological works have 
been complemented by more theoretical investigations, aiming at a QCD-based 
approach to the problem. The purpose of this talk is to distill a basic 
message from some of the latter studies~\cite{static}--\cite{nb3}. 

Let us start by defining the observables we are interested in. 
The heavy quark contribution to the spectral function of the electromagnetic
current can be written as 
\be
 \rho_V (Q) 
 \equiv 
 \int_{-\infty}^\infty 
 \!\! {\rm d}t \!
 \int \! {\rm d}^3 \vec{x}\,
 e^{i Q\cdot x}
 \left\langle
  \fr12 [ 
  \hat \mathcal{J}^\mu(x), 
  \hat \mathcal{J}_\mu(0)
   ]
 \right\rangle
 \;, \la{rhoV} 
\ee
where 
$
 \hat \mathcal{J}^\mu
 \equiv
 \,\hat{\!\bar\psi}\, \gamma^\mu\, \hat\psi  
$; 
$\hat\psi$ is the heavy quark field operator in the Heisenberg picture; 
$
 \langle \ldots \rangle \equiv {\mathcal{Z}^{-1}} 
 \Tr[ (...) e^{-\beta\hat H}]
$
is the thermal expectation value; 
$
 \beta \equiv {1}/{T} 
$
is the inverse temperature; 
and we assume the metric convention
($+$$-$$-$$-$). 
This spectral function 
determines the production rate of muon--antimuon pairs
from the system~\cite{dilepton}, 
\be
 \frac{{\rm d} N_{\mu^-\mu^+}}{{\rm d}^4 x{\rm d}^4 Q} =  
 \frac{ -2 e^4 Z^2}{3 (2\pi)^5 Q^2} 
 \biggl( 1 + \frac{2 m_\mu^2}{Q^2}
 \biggr)
 \biggl(
 1 - \frac{4 m_\mu^2}{Q^2} 
 \biggr)^\fr12 n_\rmi{\,B}(q^0) \rho_V(Q)
 \;, 
 \la{dilepton} 
\ee
where $Z$ is the heavy quark electric charge in units of $e$,
and $n_\rmi{\,B}$ is the Bose-Einstein distribution function.
In the following we assume, largely for notational simplicity, 
that the muon--antimuon pair is at rest with respect to the thermal 
medium, i.e.\ $Q \equiv (\omega, \vec{0})$.\footnote{%
 For a non-zero total spatial momentum $\vec{q}$, with $0<|\vec{q}| \ll M$, 
 the main modification would be a shift of the two-particle
 threshold, from $\omega \approx 2 M$ to 
 $\omega \approx 2 M + \vec{q}^2 / 4 M$
 (see, e.g., ref.~\cite{peskin}).
 } 
Furthermore, $M$ denotes the heavy quark (charm, bottom) pole mass. 

\begin{figure}[t]

\centerline{%
\epsfysize=6.0cm\epsfbox{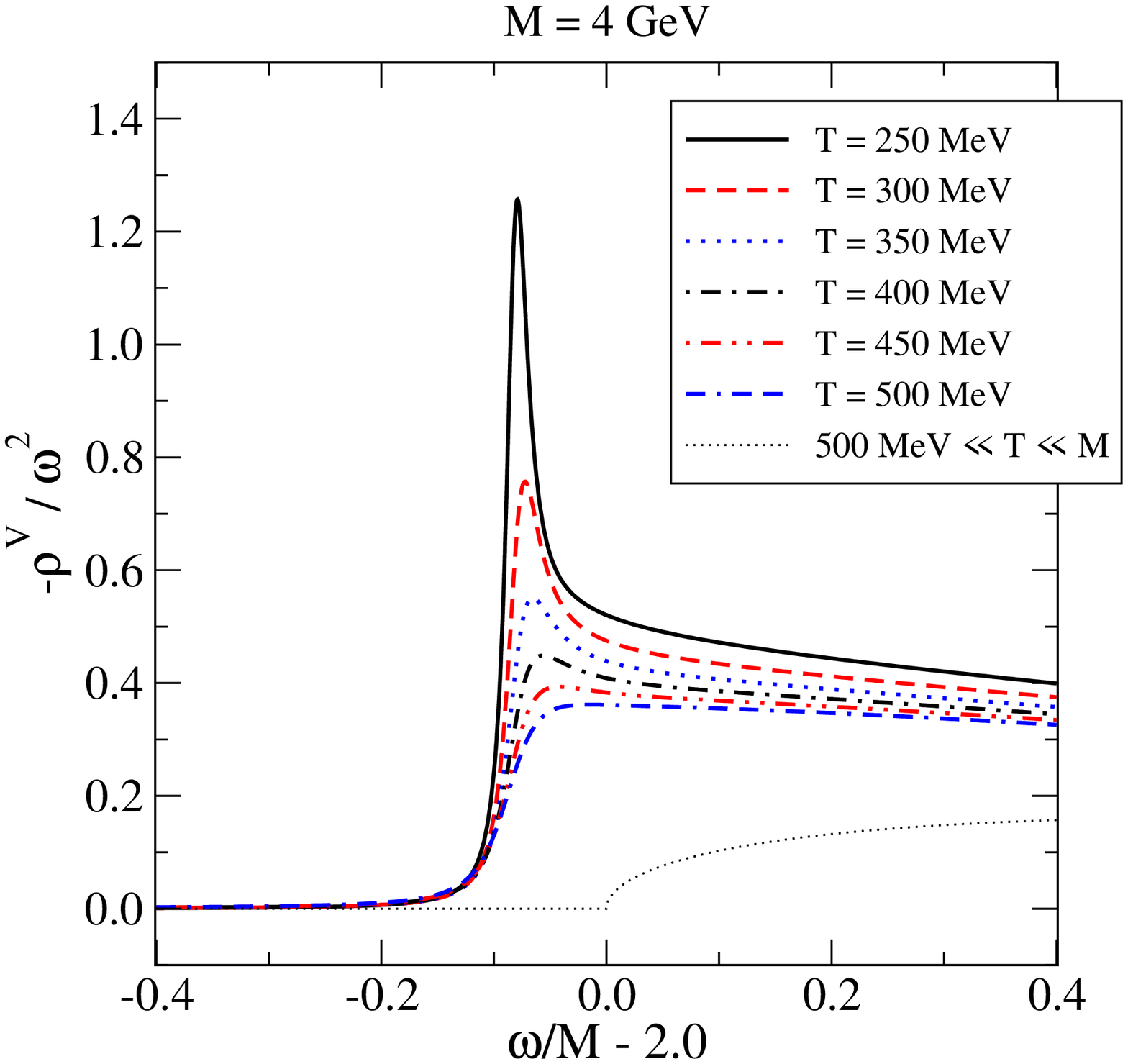}%
~~~ \epsfysize=6.0cm\epsfbox{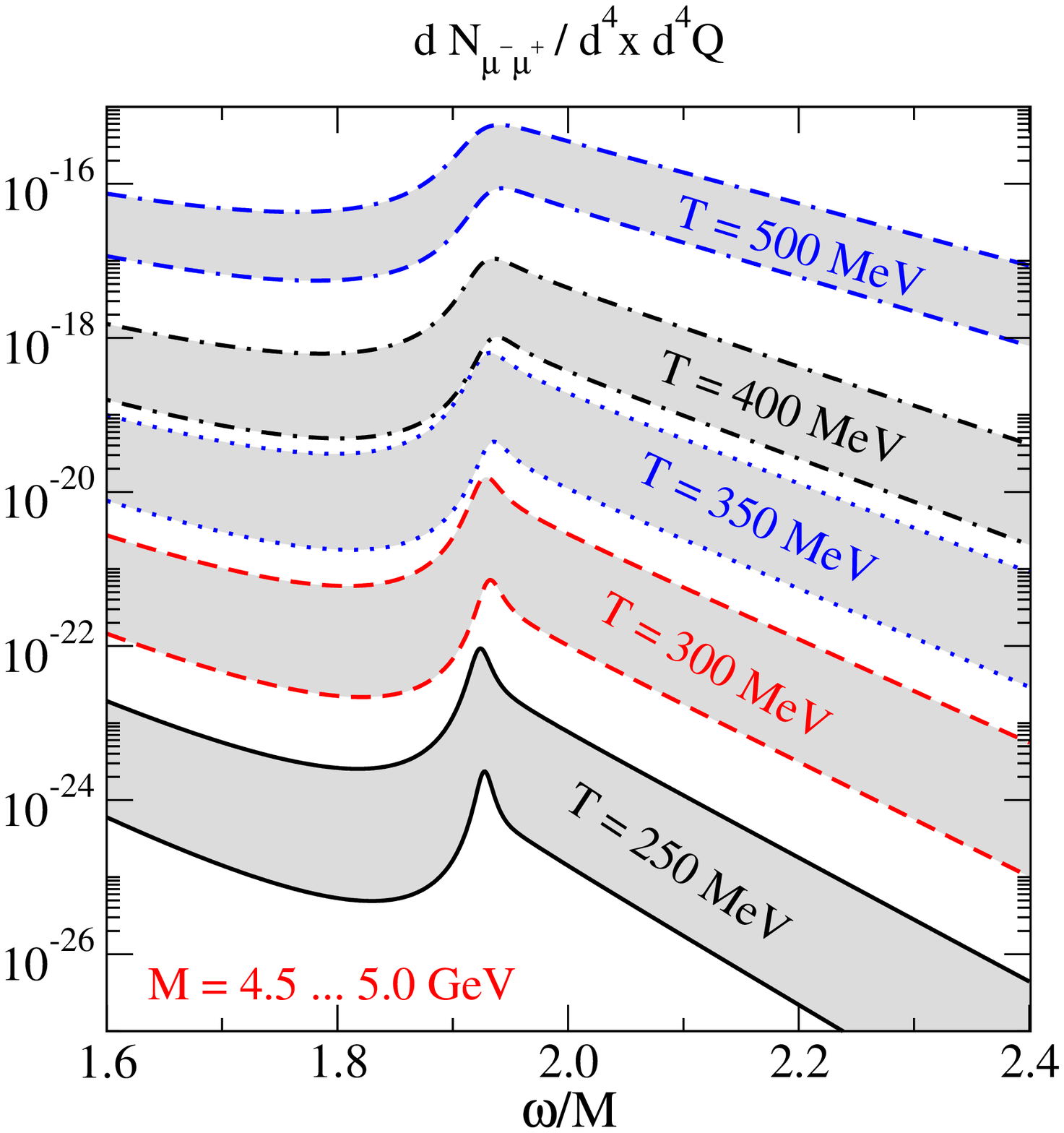}%
}

\caption[a]{Phenomenological results for (left) 
the spectral function $\rho_V$
from ref.~\cite{og2}; and (right) the dilepton
production rate 
$
 {{\rm d} N_{\mu^-\mu^+}}/{{\rm d}^4 x\,{\rm d}^4 Q} 
$
from ref.~\cite{peskin}.
In both cases $Q \equiv (\omega, \vec{0})$.
}

\la{fig:phen}
\end{figure}
  
The ultimate goal of the study would then be to compute 
$\rho_V$ and 
$
 {{\rm d} N_{\mu^-\mu^+}}/{{\rm d}^4 x\,{\rm d}^4 Q} 
$
in a certain energy 
range around the two-particle threshold, say $\omega \sim (1 \ldots 3) M$, 
and in a certain temperature range, say $T \sim (0.1 \ldots 1.0)$~GeV. 
The phenomenological interest of the problem lies in the fact that 
the observables mentioned are believed to undergo a qualitative change 
in this temperature range, thus constituting a sensitive
``thermometer'' for the system; examples of possible patterns are 
shown in \fig\ref{fig:phen}. 

Now, given that we are interested in fairly low temperatures and 
that the charm quark mass is fairly moderate compared with the QCD
scale, it would be interesting to tackle the problem with lattice 
techniques (see, e.g., refs.~\cite{latt}). Unfortunately, lattice 
methods are Euclidean, and even for a perfectly known Euclidean 
correlator $G_V(\tau)$, $\tau\in [0,\beta]$,  it is strictly speaking 
not possible to invert the relation to $\rho_V$, {\it viz.}  
\be
 G_V(\tau) = 
 \int_0^\infty
 \frac{{\rm d}\omega}{\pi} \rho_V(\omega)
 \frac{\cosh \left(\frac{\beta}{2} - \tau\right)\omega}
 {\sinh\frac{\beta \omega}{2}} 
 \;, 
\ee
without inserting further input (for instance, in momentum space, 
one must set $e^{i\beta\omega_n} \!\equiv\! 1$ 
in order to get the correct analytic continuation from 
$\omega_n$ to $\omega$). Another possible
strong coupling method, an analysis through a gravity dual 
(see, e.g., ref.~\cite{ads}), is not available for QCD.  In the 
following we therefore resort to the weak-coupling expansion, 
which at least is a theoretically consistent framework; 
in fact, particularly for the bottomonium system, experience from 
zero temperature (see, e.g., ref.~\cite{ts}) suggests that 
numerical convergence could be reasonable as well. 

%
\section{Scale hierarchy}
\label{se:scales}

A systematic weak-coupling computation can be carried out once 
the scales that enter the problem have been identified. In the heavy 
quark-mass and high-temperature limit, the QCD coupling constant $g$ 
is small, and the quarkonium states resemble non-relativistic 
systems such a positronium. Thereby we treat the heavy quark mass 
$M$ as the ``hard'' scale; the inverse Bohr radius, or relative 
momentum $M v \sim g^2 M$ as the ``soft'' scale; and the binding 
energy $M v^2 \sim g^4 M$ as the ``ultrasoft'' scale. We assume
the QCD scale to be at most as large as the ultrasoft scale.

The  energy scale with which we probe the system, $\omega$, 
could in principle be anything. However, conceivably the most 
interesting range is $|\omega - 2 M| \sim g^4M$, 
i.e. deviations from the two-particle threshold
by at most the binding energy; this is where the quarkonium 
resonance is to be found. 
In the following, we concentrate on this regime. 

Finally, we need to decide how temperature relates to 
the zero-temperature scales. Maybe quarkonium melts when 
$T \sim g^2 M$, given that then hard particles from the plasma
have enough momentum to 
kick constituents out of the bound state? However, such a temperature
cannot be high enough, since kicking happens through interactions, and those
are suppressed by $g$. On the other hand, increasing the temperature 
just a little bit, to $gT \sim g^2 M$, is certainly enough to dissociate 
the bound state, given that then even the Coulomb binding gets 
Debye screened. In fact, removing completely the Coulomb binding 
is an overkill, since a finite width 
could arise already earlier on. To summarise, the relevant range is 
somewhere between these two limits, $g^2 M < T < g M$~\cite{peskin}.
Therefore we now assume the equivalent hierarchy 
\be
 M \;\gg\; T\; > \; g^2 M \; >\; gT \;\gg \; g^4 M
 \;, \tb
\ee
and integrate out the hard scales one-by-one.

%
\section{Integrating out the scale $M$}

Integrating out the hard scale $M$ yields an effective
theory called NRQCD~\cite{cl,kt}. The topic is very well 
developed by now, and we only list some basic steps here; many 
details and references can be found, e.g., in the review~\cite{nb1}.

Starting from the QCD Lagrangian, 
$
 \mathcal{L}_\rmi{\,QCD} = \bar\psi (i \gamma^\mu D_\mu - M) \psi + 
 \mathcal{L}_\rmi{\,gluon}
$, one can first carry out a Foldy-Wouthuysen transformation:  
\be
 \bar\psi  \equiv 
 \left( \begin{array}{c} \theta \\ 
 - \phi \end{array} \right)^\dagger
 \exp\left( 
 - \frac{i \gamma^j \overleftarrow{\!D}_{\!\!j} }{2 M} \right)
 \;, \quad
 \psi \equiv
 \exp\left( \frac{i \gamma^j \overrightarrow{\!D}_{\!\!j} }{2 M} \right)
 \left( \begin{array}{c} \theta \\ \phi \end{array} \right)
 \;,
\ee 
where 
$\overleftarrow{\!D}_{\!\!j} 
 \,\equiv\, \overleftarrow{\!\partial}_{\!\!j}\! + i g A_j$, 
$\overrightarrow{\!D}_{\!\!j} 
 \,\equiv\, \overrightarrow{\!\partial}_{\!\!j}\! - i g A_j$, 
and $\theta, \phi$ are two-component spinors. 
Expanding in $1/M$ and dropping total derivatives, this yields 
\ba
 \mathcal{L}_\rmi{\,QCD} & = &   \theta^\dagger\left(
 iD_0 - M + \frac{\vec{D}^2 + \sigma\cdot g\vec{B} }{2 M} 
 \right)\theta
 + \phi^\dagger\left(
 iD_0 + M - \frac{\vec{D}^2 + \sigma\cdot g\vec{B}}{2 M}  
 \right)\phi  \nn & + &  \frac{i }{2 M} 
 \left( \theta^\dagger \sigma\cdot g\vec{E} \, \phi - 
 \phi^\dagger \sigma\cdot g\vec{E} \, \theta \right) + 
 \rmO\left(\frac{1}{M^2} \right)
 +   \mathcal{L}_\rmi{\,gluon}
 \;. \la{1oM}
\ea
The heavy masses on the first row can be shifted away
by $\theta\to e^{-iMt}\theta$, $\phi\to e^{iMt}\phi$. The mixed terms
on the second row become rapidly oscillating after this shift, 
$
 e^{2iMt} \theta^\dagger \sigma\cdot g\vec{E} \, \phi
 - e^{-2iMt} \phi^\dagger \sigma\cdot g\vec{E} \, \theta
$ , 
and describe hard reactions; such oscillations can be integrated
out yielding four-fermion operators. Finally, at the loop level, 
we need to correct for the local expansion through matching 
coefficients, which concerns both parameters, e.g.\
$ 
 M = m_{\tinymsbar}(m_{\tinymsbar}) 
  \left(1 + {g^2 C_F}/ {4\pi^2} + ... \right)
$, 
as well as composite operators, e.g.\ 
\be
 \bar\psi \gamma^k \psi = 
 \left[ \theta^\dagger \sigma_k \phi+ 
 \phi^\dagger \sigma_k \theta \right]
 \left(1 - \frac{g^2 C_F}{2\pi^2} + ... \right)
 \;. 
\ee

%
\section{Integrating out the scale $T$}

The next step is to integrate out the hard thermal scale $T$. 
In the heavy-quark sector, this does 
nothing dramatic, just corrects the parameters. In fact, 
since the thermal loops can be computed within NRQCD, all effects
are power-suppressed in mass, e.g.~\cite{dhr} 
\be
 \delta M = \frac{g^2 T^2 C_F}{12 M} 
 \;. 
\ee
In the gauge field sector, the same step yields nothing 
but the Hard Thermal Loop effective theory~\cite{htl}: 
$
 \mathcal{L}_\rmi{\,gluon} \longrightarrow
 \mathcal{L}_\rmi{\,gluon} + \mathcal{L}_\rmi{\,HTL} + ... 
$.

%
\section{Integrating out the scale $g^2M$}

The crux of the problem is the integration out of the scale $g^2M$, 
yielding an effective theory called pNRQCD~\cite{ps,nb2}. 
(To be more precise, all gluons of energy or momentum $\sim g^2 M$ 
are integrated out, while the composite quarkonium fields left over
as dynamical degrees of freedom can still have relative momentum
$\sim g^2 M$; it is only their energy that is small, $\sim g^4 M$.)
Although well 
established, this step is in many ways more delicate than the previous
ones; we only outline the basic ideas here
(for reviews, see refs.~\cite{mb1,nb1}). 

The pNRQCD setup is special in that in only applies to particular 
Green's functions; in the thermal context, we can say
that the thermal expectation value is restricted to the 
quark--antiquark sector of the Fock space, 
$\Tr[(...) \exp(-\beta\hat H)] \to 
\Tr[P_{{\bf 3^*} \otimes {\bf 3}} (...) \exp(-\beta\hat H)]$.
Thereby exponentially small effects $\sim\exp(-\beta M)$ 
are omitted. For the quark--antiquark pair, were represent colour as 
${\bf 3^*}\, \otimes\, {\bf 3} = {\bf 1}\, \oplus\, {\bf 8}$, 
and space coordinates as $\vec{x}_1 = \vec{X} + \vec{r}/2$,
$\vec{x}_2 = \vec{X} - \vec{r}/2$. The singlet field, 
$\mathrm{S}$, is diagonal in colour-space, 
$\mathrm{S} \equiv S\, \unit_{\!\Nc\times\Nc} / \sqrt{\Nc}$, 
while the octet field is traceless, 
$\mathrm{O} \equiv O^a T^a /\sqrt{T_F}$, 
where $T_F$ defines the generator normalization through
$\Tr[T^a T^b] = T_F \delta^{ab}$.
The interpolating operators for the component fields $S,O^a$
can be chosen as  
\ba
 \theta^\dagger(\vec{x}_2,t)W_{\vec{x}_2,\vec{x}_1} \phi(\vec{x}_1,t) 
 & \;\simeq\; & Z_s^{\fr12}(\vec{r}) S(\vec{X},\vec{r},t) \;, \\ 
 \theta^\dagger(\vec{x}_2,t)W_{\vec{x}_2,\vec{X}} T^a 
 W_{\vec{X},\vec{x}_1} \phi(\vec{x}_1,t) 
 & \;\simeq\; &  Z_o^{\fr12}(\vec{r}) O^a(\vec{X},\vec{r},t) \;, 
\ea
where $W$ is a straight Wilson line. Subsequently a general 
Lagrangian is written for the fields $\mathrm{S,O}$ as well as 
the ultrasoft gauge fields, respecting gauge invariance 
and expanded in $1/M$, $1/g^2 M\sim r$:  
\ba
 && \hspace*{-0.8cm} \mathcal{L} \simeq 
 \Tr\left\{
  \mathrm{S}^\dagger \!\left[ 
   i \partial_0  - V_s^{(0)} (r) + \frac{\nabla^2_\vec{r}}{M}
   -  \frac{V_s^{(1)} (r)}{M}   
 \right] \! \mathrm{S} + 
   \mathrm{O^\dagger}\! \left[ 
   i \mathcal{D}_0  - V_o^{(0)} (r)  + \frac{\nabla^2_\vec{r}}{M}
  -  \frac{V_o^{(1)} (r)}{M}   
 \right] \! \mathrm{O}
 \right. 
 \nn && \hspace*{-0.9cm} + \left.
 V_\rmii{$\!A$}(r) \left[ 
  \mathrm{O^\dagger} \vec{r} \cdot g\vec{E}\, \mathrm{S} +  
  \mathrm{S^\dagger} \vec{r} \cdot g\vec{E}\, \mathrm{O}
 \right]
 + \frac{V_\rmii{$B$}(r)}{2} \left[ 
  \mathrm{O^\dagger} \vec{r} \cdot g\vec{E}\, \mathrm{O} +  
  \mathrm{O^\dagger} \mathrm{O} \vec{r} \cdot g\vec{E} 
 \right]
 \right\}
 +  \rmO\left(r^2, \frac{1}{M^2}\right) 
 \;,
\ea
where $\mathcal{D}_0$ is the covariant derivative in the adjoint
representation, i.e.\ 
$
 i \mathcal{D}_0 \mathrm{O} = 
 i \partial_0 \mathrm{O} + g [A_0(\vec{X},t),\mathrm{O}]
$. 
As indicated, gauge fields only depend on the 
center-of-mass coordinates $\vec{X},t$, whereas dependence 
on $\vec{r}$ can be Taylor-expanded. As far as the structure 
on the second row is concerned, we remark that at leading order 
$V_A = V_B = 1$, as can be inferred from a tree-level
matching of a three-object vertex to NRQCD, on which side the 
covariant derivative can be expanded as 
$
 i D_0 \to i D_0 +  \vec{r} \cdot g\vec{E} + ... \,
$. 

Now, the effective theory contains many ``potentials'', 
$V_s^{(i)}$, $V_o^{(i)}$, $V_\rmii{$A$}$, $V_\rmii{$B$}$, $...$, 
which are to be treated as matching coefficients. 
They can be computed from various Wilson loops 
with $t \gg r$ (since $\partial_t\sim g^4 M \ll |\nabla|\sim g^2 M$), 
i.e. in the static limit. We return to the determination 
of $V_s^{(0)}$ presently. 

For the computation of the spectral function $\rho_V$, a representation
of the vector current within pNRQCD is also needed. 
The vector current is a local object,
corresponding to $\vec{r} = \vec{0}$. At zero distance, $Z_s(\vec{0})=\Nc$, 
and the relation between pNRQCD and  NRQCD  
fields becomes unambiguous. In addition, 
spin indices, which have been suppressed in the discussion above, 
can be added in a trivial way. 
Thereby the spectral function can indeed be computed within pNRQCD
(for a concise yet explicit presentation at zero temperature, 
see ref.~\cite{bkp}).

%
\section{Integrating out the scale $gT$}

Even though our ultimate goal is to treat the situation $g^2 M > gT$,
in which case the scale $gT$ is integrated out within pNRQCD, 
it will be illuminating to start by considering the case $g^2 M \sim gT$. 
Indeed the scale $gT$ can then be integrated out together with the 
scale $g^2 M$, and the problem boils down to 
determining the potentials
of the previous section in the presence of non-zero
Debye screening (since $\mD r\sim 1$). 

Luckily, this problem was considered already in ref.~\cite{static}. 
In that paper the result was called a ``real-time static potential'', 
and was in general denoted by $V_>(t,r)$; the potential $V_s^{(0)}(r)$
equals the infinite-time limit thereof,  
$V_s^{(0)}(r) = \lim_{t\to\infty}V_>(t,r)$.  
Thereby~\cite{static}
\ba
 \re V_{s}^{(0)}(r) 
  & = &  
 -\frac{g^2 C_F }{4\pi} \left[ 
 \mD + \frac{\exp(-\mD r)}{r}
 \right] 
 \;, \tb 
 \\
 \im V_{s}^{(0)}(r) 
  & = &  
 - \frac{g^2 T C_F}{4\pi} \, \phi(\mD r)
 \;, \tr 
\ea
where 
$\mD \sim gT$ is the Debye mass, 
$C_F \equiv 4/3$, and 
\be
 \phi(x) \equiv
 2 \int_0^\infty \! \frac{{\rm d} z \, z}{(z^2 +1)^2}
 \left[
   1 - \frac{\sin(z x)}{zx} 
 \right]
 \;. \tr
\ee

Let us comment, in passing, on the existence of an imaginary part
in the potential. For $r\to\infty$, the real part becomes a constant, 
corresponding to twice the thermal heavy quark mass shift~\cite{gi}; 
in complete analogy, the imaginary part at $r\to \infty$ equals~\cite{ab}
twice the known heavy quark width~\cite{rdp}. In some sense, the physics
of the imaginary part is also closely related to (single) heavy quark 
energy loss / drag force / diffusion coefficient, on which a lot
of work has been carried out recently (see, e.g., refs.~\cite{hq}): 
in fact the essential graph can be depicted in the same way, 

\begin{figure}[h]


~~\centerline{%
\raise2ex\hbox{\epsfxsize=2cm\epsfbox{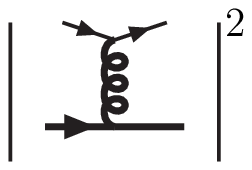}}%
\raise6ex\hbox{,}
}

\vspace*{-0.3cm}

\end{figure}

\noindent
but the integral over the spatial momentum of the gluon
is weighed differently. 

What we really wanted to do in this section, however, was to 
integrate out the scale $gT$ for $g^2 M > gT$. One possibility
would now be just to expand the previous potential in 
$\mD r \sim gT/g^2 M < 1$. It is illuminating, however, to 
really derive the result in two separate steps, like the effective
theory setup requires; the imaginary part turns out to have an $1/\epsilon$
divergence, and it is a nice crosscheck to see that this cancels 
in the end~\cite{nb3}. 

The first step consists of the computation of the potential $V_s^{(0)}$ 
but with $\mD < |\vec{p}| < T$, where $|\vec{p}|\sim 1/r \sim g^2 M$ is 
the gluon momentum. Thereby we are advised to expand the static limit
of the time-ordered HTL gluon propagator,
\be
  i D^T_{00}(0,\vec{p})
 = \frac{1}{\vec{p}^2 + \mD^2} 
 - i \frac{\pi\mD^2 T}{|\vec{p}|(\vec{p}^2 + \mD^2)^2}
 \;, 
\ee
to leading non-trivial order in $\mD^2/\vec{p}^2$, and the imaginary
part becomes
\ba
 \delta_{g^2\! M} V_{s}^{(0)}(r) & = &  
 ... +  
 i\pi g^2 C_F \mD^2 T 
 \int\! \frac{{\rm d}^{3-2\epsilon}\vec{p}}
 {(2\pi)^{3-2\epsilon}} \frac{e^{i \vec{p}\cdot\vec{r}}}{|\vec{p}|^5}
 \\  
 & = & ... +  
 i\pi g^2 C_F \mD^2 T\,  
 \frac{r^2 \mu^{-2\epsilon}}{24\pi^2}
 \biggl[
  \frac{1}{\epsilon} + \ln(r^2\bmu^2) 
  + 2 \gamma_\rmii{E} - 2 \ln 2 - 1  
 \biggr]
 \;, 
\ea
where $\bmu^2 \equiv 4\pi\mu^2 e^{-\gamma_\rmii{E}}$ 
is the scale parameter of the $\msbar$ scheme. 

In the second step, the scale $gT$ is integrated out within
the pNRQCD effective theory. The effect comes from a loop where
the singlet splits into an octet and a colour-electric field. This is
mediated by the vertex containing the potential $V_\rmii{$A$}$
(which, as mentioned, equals unity at leading order). 
For the loop momenta relevant 
for the integration, the octet propagator is $\sim \delta(p^0)$,
so again the static limit of the (colour-electric) 
gluon propagator is needed.
This leads to 
\ba
 \delta_{g T} V_{s}^{(0)}(r) & = &  
 ... - i\pi g^2 C_F \mD^2 T \,
 \fr12 \int\! \frac{{\rm d}^{3-2\epsilon}\vec{p}}
 {(2\pi)^{3-2\epsilon}} 
 \frac{(\vec{p}\cdot\vec{r})^2}{|\vec{p}| (\vec{p}^2 + \mD^2)^2 }
 \\ 
 & = & ... 
 - i\pi g^2 C_F \mD^2 T\,  
 \frac{r^2 \mu^{-2\epsilon}}{24\pi^2}
 \biggl[
  \frac{1}{\epsilon} + \ln\biggl(\frac{\bmu^2}{\mD^2}\biggr) 
  - 2 \ln 2 + \fr53  
 \biggr]
 \;. 
\ea
Summing the two parts together, we arrive at~\cite{nb3} 
\ba
 \re V_{s}^{(0)}(r) 
 & = & 
 -\frac{g^2 C_F }{4\pi} 
 \frac{1}{r} + ... 
 \;, \tb 
 \\
 \im V_{s}^{(0)}(r) 
 & = & 
 - \frac{g^2 C_F T}{4\pi} \frac{\mD^2 r^2}{3}
   \left( \ln\frac{1}{\mD r} - \gamma_\rmii{E}+\fr43 \right) + ...  
 \;. \tr 
\ea

%
\section{Physics at the scale $g^4 M$}

Having integrated out all but the ultrasoft scale, let us 
now inspect the singlet propagator at the lowest energies. 
We can expect a resonance to melt once the width appearing
in the propagator becomes as large as the binding energy. 
Of course, this condition cannot be posed in an exact way, 
and for real-world applications we would indeed be interested in 
computing $\rho_V$ in some macroscopic energy and temperature range; 
nevertheless, for the purposes of this talk, we restrict to 
the task of estimating the melting temperature.

Parametrically (omitting $C_F/4\pi$ and numerical factors), 
the equality of the binding energy and width from the end
of the previous section reads
\be
 \frac{g^2}{r} \sim g^2 T \mD^2 r^2 \ln \frac{1}{\mD r} \;. 
\ee
Inserting $r\sim 1/g^2 M$ and $\mD \sim gT$, we get
\be
 g^4 M^3 \sim T^3 \ln \frac{g M}{T} \;. 
\ee
For small enough $g$ this can be solved in a leading-logarithmic
approximation, yielding
\be
 T \sim g^{\fr43} (\ln{\textstyle\frac{1}{g}} )^{-\fr13} M \;. 
\ee
This result was first obtained (without the logarithm)
in ref.~\cite{es}, and it indeed lies within the range
$g^2 M < T < g M$ discussed in \se\ref{se:scales}, 
as must be the case~\cite{peskin}.

%
\section{Conclusions}

The main purpose of this talk has been to underline 
the question posed by the title. Indeed, it appears that quarkonium 
physics at high temperatures is important enough phenomenologically
to deserve some theoretically-minded consideration as well. 

As far as an answer to the question is concerned, 
we have tried to obtain one 
within the framework
of the weak-coupling expansion. The most important point to 
realise is that a hierarchy can be found between the different
dynamical scales affecting the problem. In fact, supplementing
the discussion so far with the non-perturbative colour-magnetic
scale $g^2 T$, the hierarchy relevant for the melting of 
quarkonium can be written as 
\be
 M \;\gg\; T\; > \; g^2 M \; >\; gT \; >\; g^2T \; > \; g^4 M
 \;.
\ee

Furthermore, for energies around the two-particle threshold, 
$|\omega - 2 M|\sim g^4 M$, we have argued that the way to exploit
the scale hierarchy is to generalize the effective theories known
as NRQCD and pNRQCD to finite temperatures~\cite{es,nb3}. 
As a result of 
such an analysis, it can be argued that quarkonium melts at 
$T \sim g^{4/3} M$~\cite{es}. We find it comforting that the 
phenomenon of melting can thus be confirmed model-independently. 

For practical applications of the effective theory setup, 
the issue of the convergence of the weak-coupling series must
be addressed. For this purpose, it would be helpful to use
the framework to compute higher order corrections to the quarkonium 
spectral function. Also, the influence of the colour-magnetic scale $g^2T$ 
needs to be estimated~\cite{imV}. It appears reasonable to expect
that as a result of such work, at least the study of the bottomonium 
resonance can ultimately be promoted to a quantitative level. 

Finally, irrespective of the numerical convergence of 
the weak-coupling expansion, it is perhaps useful to emphasize
the new qualitative features that this 
approach has unveiled. In particular, the fact that the real-time static 
potential has an unsuppressed $\vec{r}$-dependent imaginary part
at finite temperatures should in our opinion
be taken into account also in phenomenological 
potential model studies.

%

\vspace*{4mm}

\noindent {\bf Acknowledgements}

\vspace*{3mm}


I wish to thank the University of Oulu, Finland, 
where this talk was prepared, for kind hospitality. 
The work was partly supported by the BMBF project
{\em Hot Nuclear Matter from Heavy Ion Collisions 
     and its Understanding from QCD}, 
and by the Academy of Finland, contract no.\ 122079.



\end{document}